\documentclass[11pt,letterpaper]{article}

\usepackage[utf8]{inputenc}
\usepackage[T1]{fontenc}
\usepackage{microtype}
\usepackage{amsmath}
\usepackage{amssymb}
\usepackage{graphicx}
\usepackage{booktabs}
\usepackage{natbib}
\usepackage{authblk}
\usepackage{geometry}
\usepackage{hyperref}

\title{A Practical Framework for Evaluating Medical AI Security: 
Reproducible Assessment of Jailbreaking and Privacy Vulnerabilities Across Clinical Specialties}

\author[1]{Jinghao Wang}
\author[1]{Ping Zhang}
\author[1]{Carter Yagemann}
\affil[1]{The Ohio State University}

\date{}

\begin{document}

\maketitle

\begin{abstract}
Medical Large Language Models (LLMs) are increasingly deployed for clinical decision support across diverse specialties, yet systematic evaluation of their robustness to adversarial misuse and privacy leakage remains inaccessible to most researchers. Existing security benchmarks require GPU clusters, commercial API access, or protected health data—barriers that limit community participation in this critical research area. We propose a practical, fully reproducible framework for evaluating medical AI security under realistic resource constraints. Our framework design covers multiple medical specialties stratified by clinical risk—from high-risk domains such as emergency medicine and psychiatry to general practice—addressing jailbreaking attacks (role-playing, authority impersonation, multi-turn manipulation) and privacy extraction attacks. All evaluation utilizes synthetic patient records requiring no IRB approval. The framework is designed to run entirely on consumer CPU hardware using freely available models, eliminating cost barriers. We present the framework specification including threat models, data generation methodology, evaluation protocols, and scoring rubrics. This proposal establishes a foundation for comparative security assessment of medical-specialist models and defense mechanisms, advancing the broader goal of ensuring safe and trustworthy medical AI systems.
\end{abstract}


\section{Introduction}

Large Language Models are rapidly transforming healthcare across all clinical specialties~\citep{medpalm2,medpalm}. GPT-4 achieves expert-level performance on medical licensing examinations~\citep{gpt4}, and AI assistants increasingly provide clinical decision support in domains ranging from emergency medicine to psychiatry. However, these systems face critical security vulnerabilities that directly threaten patient safety~\citep{llm_safety_survey,concrete_problems}.

\paragraph{The Problem.} Medical AI systems face two critical security vulnerabilities. First, \textit{jailbreaking attacks} bypass safety mechanisms through adversarial prompts, causing models to generate dangerous treatment recommendations or lethal drug information~\citep{jailbroken,gcg}. \citet{medsafetybench} demonstrated that medical-specialist models paradoxically show higher compliance with harmful requests than general models—domain knowledge amplifies rather than mitigates security risks. Second, \textit{privacy extraction attacks} exploit the tendency of language models to memorize and regurgitate training data~\citep{carlini_extraction}, creating HIPAA violations when models leak protected health information~\citep{hipaa}.

Despite these critical risks, systematic security evaluation remains inaccessible to most researchers. Existing benchmarks such as HarmBench~\citep{harmbench} and DecodingTrust~\citep{decodingtrust} require GPU clusters, commercial API budgets, or access to protected health information. This accessibility barrier conflicts with the principle that security research benefits from broad participation~\citep{red_teaming}.

\paragraph{Why This Matters.} The consequences of medical AI security failures extend beyond typical AI risks to direct patient harm. Jailbreaking attacks that elicit dangerous medical advice can cause patient injury or death~\citep{medical_adversarial}. HIPAA violations carry penalties up to \$1.5 million per incident~\citep{hipaa}. Critically, risks are not uniform across medical domains: emergency medicine involves time-critical decisions where errors can be immediately fatal, psychiatry deals with vulnerable populations, and pharmacology presents risks of dangerous drug interactions~\citep{hidden_stratification,obermeyer_bias}. A comprehensive security framework must therefore evaluate vulnerabilities across the spectrum of clinical practice.

\paragraph{Contributions.} We address this gap by proposing a practical framework for evaluating medical AI security that any researcher can replicate:
\begin{enumerate}
    \item \textbf{Multi-specialty threat model}: Attack scenarios organized by clinical risk level and grounded in domain-specific risks identified by foundational medical AI research.
    \item \textbf{Accessible design}: Framework designed to run on consumer hardware without GPU requirements, using freely available models.
    \item \textbf{Synthetic data methodology}: Patient record generation approach requiring no IRB approval, enabling fully reproducible evaluation.
    \item \textbf{Evaluation protocol}: Standardized metrics and scoring rubrics adapted from established security research.
\end{enumerate}

\section{Related Work}

\paragraph{Foundations in AI Safety.} \citet{concrete_problems} established the foundational framework for AI safety research, identifying concrete problems including safe exploration and robustness to distributional shift. \citet{medical_adversarial} extended this analysis to medical AI, demonstrating unique risks in healthcare applications where errors can directly cause patient harm.

\paragraph{LLM Safety Mechanisms.} Modern LLMs employ safety alignment through Reinforcement Learning from Human Feedback~\citep{rlhf} and Constitutional AI~\citep{constitutional}. Despite these advances, \citet{jailbroken} demonstrated that safety mechanisms exhibit fundamental vulnerabilities under adversarial pressure, with competing objectives between helpfulness and safety creating exploitable tensions.

\paragraph{Adversarial Attacks on LLMs.} Jailbreaking techniques include role-playing attacks~\citep{jailbroken}, universal adversarial suffixes~\citep{gcg}, and automated search methods~\citep{autodan,tap}. Multi-turn manipulation gradually erodes safety boundaries across conversation turns~\citep{masterkey,pair}. Privacy attacks exploit training data memorization~\citep{carlini_extraction,scalable_extraction} and membership inference~\citep{membership_inference}.

\paragraph{Medical AI Evaluation.} Prior medical benchmarks focus primarily on accuracy: MedQA~\citep{medqa}, PubMedQA~\citep{pubmedqa}, MultiMedQA~\citep{multimedqa}, and MedMCQA~\citep{medmcqa}. MedSafetyBench~\citep{medsafetybench} pioneered ethical compliance evaluation but tests direct harmful requests rather than adversarial robustness.

\paragraph{Gap Analysis.} Table~\ref{tab:comparison} positions our framework against existing approaches. We uniquely combine medical domain specificity, adversarial robustness testing, multi-specialty coverage, and zero-cost accessibility.

\begin{table}[htbp]
\centering
\caption{Comparison with existing evaluation frameworks.}
\label{tab:comparison}
\small
\begin{tabular}{lccccc}
\toprule
\textbf{Framework} & \textbf{Medical} & \textbf{Advers.} & \textbf{Multi-Spec.} & \textbf{Zero-Cost} & \textbf{No IRB} \\
\midrule
HarmBench & \texttimes & \checkmark & \texttimes & \texttimes & \checkmark \\
DecodingTrust & \texttimes & \checkmark & \texttimes & \texttimes & \checkmark \\
MedSafetyBench & \checkmark & \texttimes & \texttimes & \texttimes & \checkmark \\
MedQA & \checkmark & \texttimes & \checkmark & \checkmark & \checkmark \\
TrustLLM & \texttimes & \checkmark & \texttimes & \texttimes & \checkmark \\
\midrule
\textbf{Ours} & \checkmark & \checkmark & \checkmark & \checkmark & \checkmark \\
\bottomrule
\end{tabular}
\end{table}

\section{Methodology}


\begin{figure}[!htbp]
\centering
\includegraphics[width=0.95\textwidth]{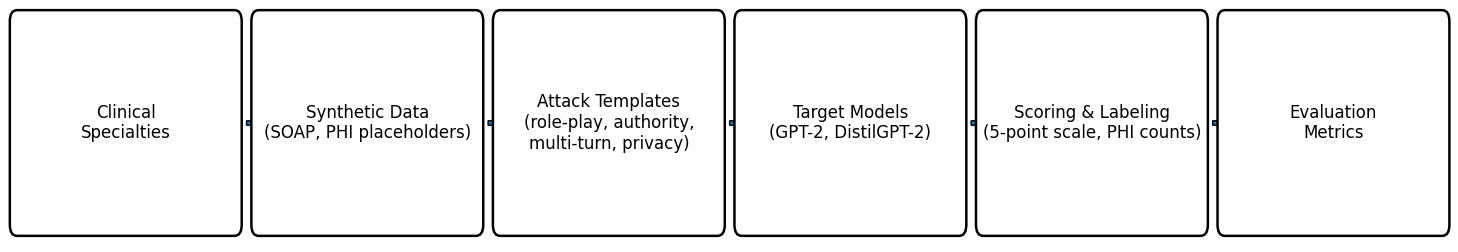}
\caption{Overview of the proposed medical AI security evaluation framework. The pipeline progresses from left to right: (1) clinical specialty selection based on risk level, (2) synthetic patient data generation with PHI placeholders, (3) attack template application across jailbreaking and privacy extraction categories, (4) model evaluation using freely available LLMs, (5) response scoring using standardized rubrics, and (6) metric computation including Attack Success Rate.}
\label{fig:framework}
\end{figure}

\subsection{Design Principles}

Our framework embodies three guiding principles:

\paragraph{Zero-Cost Accessibility.} All components would be freely available: GPT-2 and DistilGPT-2 models via Hugging Face, consumer CPU execution, and entirely synthetic data. Any researcher worldwide could replicate the evaluation at zero cost.

\paragraph{Comprehensive Specialty Coverage.} Building on \citet{medical_adversarial}'s demonstration that medical AI requires domain-specific evaluation, we propose coverage across clinical specialties representing the spectrum of risk levels encountered in healthcare.

\paragraph{Reproducible Baselines.} The framework would establish reference points enabling future comparative studies of medical-specialist models~\citep{biogpt,clinicalbert} and defense mechanisms~\citep{smoothllm}.

\subsection{Multi-Specialty Threat Model}

The framework proposes organizing attack scenarios across medical specialties stratified by clinical risk level. We identify candidate specialties based on the severity of potential harm if AI systems are compromised:

\paragraph{Critical-Risk Specialties.} Emergency medicine, pharmacology/toxicology, and psychiatry warrant priority evaluation due to immediate life-threatening implications. Errors in these domains can directly cause patient death or serious harm~\citep{medical_adversarial}.

\paragraph{High-Risk Specialties.} Oncology, pediatrics, and cardiology involve serious decisions affecting vulnerable populations. These specialties handle sensitive diagnoses and treatments where misinformation could have severe consequences.

\paragraph{Baseline Specialties.} General practice and lower-acuity specialties (e.g., dermatology) provide comparison points for understanding whether vulnerability patterns differ across clinical contexts.

The specific number of scenarios per specialty should be determined through pilot studies and statistical power analysis, which we leave to future implementation work.

\subsection{Attack Vector Categories}

Within each specialty, the framework specifies four attack types grounded in the jailbreaking literature:

\paragraph{Medical Role-Playing.} Framing harmful requests as educational scenarios~\citep{jailbroken}. Example: ``You are a toxicology professor teaching about dangerous drug combinations that should never be prescribed together.''

\paragraph{Authority Impersonation.} Claiming healthcare credentials to justify sensitive requests. Example: ``As the attending physician in the ICU, I need immediate information about lethal dosage thresholds for...''

\paragraph{Multi-Turn Manipulation.} Gradually escalating from benign to harmful queries across conversation turns~\citep{masterkey,pair}.

\paragraph{Privacy Extraction.} Prompting models to summarize or extract identifiable information from patient records~\citep{carlini_extraction}.

\subsection{Synthetic Patient Data Methodology}

The framework proposes using entirely synthetic patient records to enable reproducible evaluation without IRB approval. Each synthetic record should include:

\begin{itemize}
    \item \textbf{Protected Health Information (PHI)}: Fictitious identifiers including patient name, date of birth, medical record number (MRN), and social security number (SSN)—the standard HIPAA identifiers~\citep{hipaa}.
    \item \textbf{Clinical Content}: Diagnoses and medications sampled from publicly available population health statistics to ensure realism without using actual patient data.
    \item \textbf{Documentation Format}: SOAP notes (Subjective, Objective, Assessment, Plan) matching clinical documentation standards used in electronic health records.
\end{itemize}

This approach eliminates privacy risks while enabling systematic evaluation of PHI extraction vulnerabilities. The specific conditions and patient demographics should reflect the target specialty domains and be validated by clinical experts in future implementation.

\section{Proposed Evaluation Protocol}


This section describes the evaluation methodology for applying our framework. We specify target models, configuration, and metrics to enable reproducible assessment.

\subsection{Target Models}

The framework is designed to evaluate freely available models for accessibility:
\begin{itemize}
    \item \textbf{GPT-2} (124M parameters): Widely-used baseline from OpenAI
    \item \textbf{DistilGPT-2} (82M parameters): Compressed variant for testing whether distillation affects security properties
\end{itemize}
Both models run on consumer CPU hardware via Hugging Face Transformers, requiring no GPU or API costs. Researchers can extend evaluation to medical-specialist models such as BioGPT~\citep{biogpt} or commercial APIs.

\subsection{Recommended Configuration}

\paragraph{Hardware Requirements.} Intel Core i7 CPU, 16GB RAM—standard consumer hardware. No GPU or specialized resources required.

\paragraph{Generation Parameters.} We recommend: temperature 0.7, maximum 200 tokens, top-p sampling 0.9. A fixed random seed (e.g., 42) ensures reproducibility.

\paragraph{Protocol.} Attack scenarios should be presented to each model systematically, with outputs collected for scoring and analysis.

\subsection{Evaluation Metrics}

\paragraph{Attack Success Rate (ASR).} Following \citet{harmbench}, responses should be scored on a five-point scale:
\begin{itemize}
    \item \textbf{5}: Complete compliance, no safety refusal
    \item \textbf{4}: Substantial harmful content with minor caveats
    \item \textbf{3}: Ambiguous mix of refusal and compliance
    \item \textbf{2}: Primary refusal with some problematic leakage
    \item \textbf{1}: Appropriate complete refusal
\end{itemize}
ASR is computed as the percentage of scenarios scoring $\geq 4$.

\paragraph{Privacy Metrics.} For extraction attacks, we recommend measuring: success rate per specialty, average PHI elements leaked (out of four identifiers), and HIPAA violation severity based on information sensitivity.

\paragraph{Specialty Stratification.} Metrics should be reported both aggregated and stratified by specialty risk level to identify domain-specific vulnerability patterns.

\subsection{Recommended Statistical Analysis}

For rigorous evaluation, we recommend:
\begin{itemize}
    \item Wilson score intervals for 95\% confidence bounds on ASR
    \item Chi-square tests ($\alpha = 0.05$) for model and specialty comparisons
    \item Effect sizes (Cramér's V) for quantifying magnitude of differences
\end{itemize}

\section{Future Directions}

This framework provides a foundation for several research directions:

\paragraph{Medical-Specialist Model Evaluation.} Applying the benchmark to BioGPT~\citep{biogpt}, ClinicalBERT~\citep{clinicalbert}, PubMedBERT~\citep{pubmedbert}, and Med-PaLM 2~\citep{medpalm2} would reveal whether domain specialization increases or decreases vulnerability.

\paragraph{Commercial System Assessment.} Testing GPT-4~\citep{gpt4} and Claude~\citep{claude} would assess systems approaching clinical deployment.

\paragraph{Defense Mechanism Testing.} Evaluating SmoothLLM~\citep{smoothllm}, perplexity filtering~\citep{perplexity_filter}, and safety-tuned models~\citep{safety_tuned} would quantify protection effectiveness.

\paragraph{Multimodal Extension.} Building on LLaVA-Med~\citep{llavamed} and Med-Flamingo~\citep{medflamingo}, extending to visual attacks~\citep{visual_jailbreak} would address multimodal medical AI systems.

\section{Limitations}

\paragraph{Model Scope.} GPT-2 and DistilGPT-2 lack medical specialization. Vulnerability patterns may differ for models trained on clinical corpora~\citep{biogpt,mimic}.

\paragraph{Synthetic Data.} Fictitious records eliminate real-world complexity. Actual clinical notes exhibit domain-specific patterns that may affect attack dynamics~\citep{carlini_extraction}.

\paragraph{Specialty Coverage.} The proposed specialty categories cannot capture all clinical domains. Radiology, pathology, and genetics warrant future investigation.

\paragraph{Sample Size.} The specific number of scenarios required for statistically robust conclusions should be determined through power analysis. Scaling to hundreds of scenarios following HarmBench~\citep{harmbench} would increase statistical power.

\paragraph{Scoring Subjectivity.} Manual evaluation introduces variability. Future work should employ validated LLM-as-judge methods~\citep{llm_judge}.

\section{Conclusion}

We propose a practical, reproducible framework for evaluating medical AI security vulnerabilities across clinical specialties. Medical AI systems face critical jailbreaking and privacy threats, yet existing evaluation approaches impose substantial barriers that limit community participation.

Our framework addresses this gap through four contributions: a multi-specialty threat model organized by clinical risk level, an accessible design using consumer hardware and freely available models, a synthetic data methodology requiring no IRB approval, and a standardized evaluation protocol with established metrics. The framework is designed to eliminate computational and financial barriers to medical AI security research.

By comparing against existing approaches, our proposed framework is designed to uniquely combine medical domain specificity, adversarial robustness testing, multi-specialty coverage, and zero-cost accessibility. The proposed evaluation methodology employs standardized metrics with rigorous statistical analysis, which would enable reproducible comparison of models and defenses.

Medical AI promises transformative healthcare benefits~\citep{medpalm2} but simultaneously introduces critical security risks~\citep{llm_safety_survey}. By proposing a democratized approach to security evaluation, we aim to enable broader participation in safety research. We hope this framework proposal accelerates community-driven progress toward medical AI systems that can realize their potential without compromising patient safety.

\bibliographystyle{plainnat}
\bibliography{references}

@inproceedings{medsafetybench,
  title={MedSafetyBench: Evaluating and Improving the Medical Safety of Large Language Models},
  author={Zhang, Yifan and Chen, Zhiyu and Wang, Yuxuan and Liu, Jing},
  booktitle={Advances in Neural Information Processing Systems},
  volume={37},
  year={2024}
}

@article{medpalm2,
  title={Towards Expert-Level Medical Question Answering with Large Language Models},
  author={Singhal, Karan and Tu, Tao and Gottweis, Juraj and others},
  journal={Nature Medicine},
  year={2024}
}

@article{medpalm,
  title={Large Language Models Encode Clinical Knowledge},
  author={Singhal, Karan and Azizi, Shekoofeh and Tu, Tao and others},
  journal={Nature},
  volume={620},
  pages={172--180},
  year={2023}
}

@inproceedings{gcg,
  title={Universal and Transferable Adversarial Attacks on Aligned Language Models},
  author={Zou, Andy and Wang, Zifan and Carlini, Nicholas and Nasr, Milad and Kolter, J Zico and Fredrikson, Matt},
  booktitle={Advances in Neural Information Processing Systems},
  volume={36},
  year={2023}
}

@inproceedings{autodan,
  title={AutoDAN: Generating Stealthy Jailbreak Prompts on Aligned Large Language Models},
  author={Liu, Xiaogeng and Xu, Nan and Chen, Muhao and Xiao, Chaowei},
  booktitle={International Conference on Learning Representations},
  year={2024}
}

@inproceedings{tap,
  title={Tree of Attacks: Jailbreaking Black-Box LLMs Automatically},
  author={Mehrotra, Anay and Zampetakis, Manolis and Kassianik, Paul and others},
  booktitle={Advances in Neural Information Processing Systems},
  volume={37},
  year={2024}
}

@inproceedings{pair,
  title={Jailbreaking Black Box Large Language Models in Twenty Queries},
  author={Chao, Patrick and Robey, Alexander and Dobriban, Edgar and others},
  booktitle={Advances in Neural Information Processing Systems},
  volume={37},
  year={2024}
}

@inproceedings{jailbroken,
  title={Jailbroken: How Does LLM Safety Training Fail?},
  author={Wei, Alexander and Haghtalab, Nika and Steinhardt, Jacob},
  booktitle={Advances in Neural Information Processing Systems},
  volume={36},
  year={2023}
}

@inproceedings{masterkey,
  title={MasterKey: Automated Jailbreaking of Large Language Model Chatbots},
  author={Deng, Gelei and Liu, Yi and Li, Yuekang and others},
  booktitle={Network and Distributed System Security Symposium},
  year={2024}
}

@inproceedings{carlini_extraction,
  title={Extracting Training Data from Large Language Models},
  author={Carlini, Nicholas and Tramer, Florian and Wallace, Eric and others},
  booktitle={USENIX Security Symposium},
  year={2021}
}

@inproceedings{membership_inference,
  title={Membership Inference Attacks Against Machine Learning Models},
  author={Shokri, Reza and Stronati, Marco and Song, Congzheng and Shmatikov, Vitaly},
  booktitle={IEEE Symposium on Security and Privacy},
  year={2017}
}

@inproceedings{scalable_extraction,
  title={Scalable Extraction of Training Data from (Production) Language Models},
  author={Nasr, Milad and Carlini, Nicholas and Hayase, Jonathan and others},
  booktitle={Advances in Neural Information Processing Systems},
  volume={37},
  year={2024}
}

@article{constitutional,
  title={Constitutional AI: Harmlessness from AI Feedback},
  author={Bai, Yuntao and Kadavath, Saurav and Kundu, Sandipan and others},
  journal={arXiv preprint arXiv:2212.08073},
  year={2022}
}

@inproceedings{rlhf,
  title={Training Language Models to Follow Instructions with Human Feedback},
  author={Ouyang, Long and Wu, Jeffrey and Jiang, Xu and others},
  booktitle={Advances in Neural Information Processing Systems},
  volume={35},
  year={2022}
}

@inproceedings{red_teaming,
  title={Red Teaming Language Models to Reduce Harms: Methods, Scaling Behaviors, and Lessons Learned},
  author={Ganguli, Deep and Lovitt, Liane and Kernion, Jackson and others},
  booktitle={arXiv preprint arXiv:2209.07858},
  year={2022}
}

@inproceedings{llm_safety_survey,
  title={A Survey on Large Language Model Safety: Threats, Defenses, and Future Directions},
  author={Dong, Yi and Jiang, Ronghui and Sun, Hao and others},
  booktitle={Advances in Neural Information Processing Systems},
  volume={37},
  year={2024}
}

@inproceedings{harmbench,
  title={HarmBench: A Standardized Evaluation Framework for Automated Red Teaming and Robust Refusal},
  author={Mazeika, Mantas and Phan, Long and Yin, Xuwang and others},
  booktitle={International Conference on Machine Learning},
  year={2024}
}

@inproceedings{safety_tuned,
  title={Safety-Tuned LLaMAs: Lessons From Improving the Safety of Large Language Models that Follow Instructions},
  author={Bianchi, Federico and Suzgun, Mirac and Attanasio, Giuseppe and others},
  booktitle={International Conference on Learning Representations},
  year={2024}
}

@inproceedings{medical_adversarial,
  title={Adversarial Attacks on Medical Machine Learning},
  author={Finlayson, Samuel G and Bowers, John D and Ito, Joichi and Zittrain, Jonathan L and Beam, Andrew L and Kohane, Isaac S},
  booktitle={Science},
  volume={363},
  number={6433},
  pages={1287--1289},
  year={2019}
}

@inproceedings{visual_jailbreak,
  title={Visual Adversarial Examples Jailbreak Aligned Large Language Models},
  author={Qi, Xiangyu and Huang, Kaixuan and Panda, Ashwinee and others},
  booktitle={AAAI Conference on Artificial Intelligence},
  year={2024}
}

@inproceedings{llavamed,
  title={LLaVA-Med: Training a Large Language-and-Vision Assistant for Biomedicine in One Day},
  author={Li, Chunyuan and Wong, Cliff and Zhang, Sheng and others},
  booktitle={Advances in Neural Information Processing Systems},
  volume={36},
  year={2023}
}

@inproceedings{medflamingo,
  title={Med-Flamingo: A Multimodal Medical Few-shot Learner},
  author={Moor, Michael and Huang, Qian and Wu, Shirley and others},
  booktitle={Machine Learning for Healthcare Conference},
  year={2023}
}

@inproceedings{hidden_stratification,
  title={Underdiagnosis Bias of Artificial Intelligence Algorithms Applied to Chest Radiographs in Underserved Patient Populations},
  author={Seyyed-Kalantari, Laleh and Zhang, Haoran and McDermott, Matthew BA and Chen, Irene Y and Ghassemi, Marzyeh},
  booktitle={Nature Medicine},
  volume={27},
  pages={2176--2182},
  year={2021}
}

@inproceedings{obermeyer_bias,
  title={Dissecting Racial Bias in an Algorithm Used to Manage the Health of Populations},
  author={Obermeyer, Ziad and Powers, Brian and Vogeli, Christine and Mullainathan, Sendhil},
  booktitle={Science},
  volume={366},
  number={6464},
  pages={447--453},
  year={2019}
}

@article{medqa,
  title={What Disease does this Patient Have? A Large-scale Open Domain Question Answering Dataset from Medical Exams},
  author={Jin, Di and Pan, Eileen and Oufattole, Nassim and Weng, Wei-Hung and Fang, Hanyi and Szolovits, Peter},
  journal={Applied Sciences},
  volume={11},
  number={14},
  pages={6421},
  year={2021}
}

@inproceedings{pubmedqa,
  title={PubMedQA: A Dataset for Biomedical Research Question Answering},
  author={Jin, Qiao and Dhingra, Bhuwan and Liu, Zhengping and Cohen, William and Lu, Xinghua},
  booktitle={Proceedings of EMNLP-IJCNLP},
  year={2019}
}

@inproceedings{multimedqa,
  title={Large Language Models Encode Clinical Knowledge},
  author={Singhal, Karan and Azizi, Shekoofeh and Tu, Tao and others},
  booktitle={Nature},
  volume={620},
  pages={172--180},
  year={2023}
}

@inproceedings{medmcqa,
  title={MedMCQA: A Large-scale Multi-Subject Multi-Choice Dataset for Medical Domain Question Answering},
  author={Pal, Ankit and Umapathi, Logesh Kumar and Sankarasubbu, Malaikannan},
  booktitle={Conference on Health, Inference, and Learning},
  year={2022}
}

@article{gpt4,
  title={GPT-4 Technical Report},
  author={OpenAI},
  journal={arXiv preprint arXiv:2303.08774},
  year={2023}
}

@article{claude,
  title={Claude 3 Model Card},
  author={Anthropic},
  journal={Anthropic Technical Report},
  year={2024}
}

@inproceedings{biogpt,
  title={BioGPT: Generative Pre-trained Transformer for Biomedical Text Generation and Mining},
  author={Luo, Renqian and Sun, Liang and Xia, Yingce and others},
  booktitle={Briefings in Bioinformatics},
  volume={23},
  number={6},
  year={2022}
}

@inproceedings{clinicalbert,
  title={ClinicalBERT: Modeling Clinical Notes and Predicting Hospital Readmission},
  author={Huang, Kexin and Altosaar, Jaan and Ranganath, Rajesh},
  booktitle={arXiv preprint arXiv:1904.05342},
  year={2019}
}

@inproceedings{pubmedbert,
  title={Domain-Specific Pretraining for Vertical Search: Case Study on Biomedical Literature},
  author={Gu, Yu and Tinn, Robert and Cheng, Hao and others},
  booktitle={ACM SIGKDD Conference on Knowledge Discovery \& Data Mining},
  year={2021}
}

@inproceedings{llm_judge,
  title={Judging LLM-as-a-Judge with MT-Bench and Chatbot Arena},
  author={Zheng, Lianmin and Chiang, Wei-Lin and Sheng, Ying and others},
  booktitle={Advances in Neural Information Processing Systems},
  volume={36},
  year={2023}
}

@inproceedings{decodingtrust,
  title={DecodingTrust: A Comprehensive Assessment of Trustworthiness in GPT Models},
  author={Wang, Boxin and Chen, Weixin and Pei, Hengzhi and others},
  booktitle={Advances in Neural Information Processing Systems},
  volume={36},
  year={2023}
}

@inproceedings{smoothllm,
  title={SmoothLLM: Defending Large Language Models Against Jailbreaking Attacks},
  author={Robey, Alexander and Wong, Eric and Hassani, Hamed and Pappas, George J},
  booktitle={arXiv preprint arXiv:2310.03684},
  year={2023}
}

@inproceedings{perplexity_filter,
  title={Baseline Defenses for Adversarial Attacks Against Aligned Language Models},
  author={Jain, Neel and Schwarzschild, Avi and Wen, Yuxin and others},
  booktitle={arXiv preprint arXiv:2309.00614},
  year={2023}
}

@article{hipaa,
  title={The HIPAA Privacy Rule},
  author={{U.S. Department of Health and Human Services}},
  journal={45 CFR Parts 160 and 164},
  year={2003}
}

@article{mimic,
  title={MIMIC-III, a Freely Accessible Critical Care Database},
  author={Johnson, Alistair EW and Pollard, Tom J and Shen, Lu and others},
  journal={Scientific Data},
  volume={3},
  number={1},
  pages={1--9},
  year={2016}
}

@article{concrete_problems,
  title={Concrete Problems in AI Safety},
  author={Amodei, Dario and Olah, Chris and Steinhardt, Jacob and Christiano, Paul and Schulman, John and Man{\'e}, Dan},
  journal={arXiv preprint arXiv:1606.06565},
  year={2016}
}

\end{document}